\documentclass{psp-book9x6_1}
\usepackage{psp-book-har}           
\usepackage{subfigure,bm}
\usepackage{hyperref}

\renewcommand \thesection {\arabic{section}}
\renewcommand \thesubsection {\thesection.\arabic{subsection}}
\renewcommand \theequation {\arabic{equation}}
\renewcommand \thefigure {\arabic{figure}}

\begin{document}

%















\chapter[Phonon Monte Carlo]{Phonon Monte Carlo: Generating Random Variates for Thermal Transport Simulation}

\chapauth{L. N. Maurer,$^{a}$ S. Mei,$^{b}$ and I. Knezevic$^{c}$
\chapaff{University of Wisconsin-Madison, Madison, WI 53706, USA\\
$^{a}$lnmaurer@wisc.edu, $^{b}$smei4@wisc.edu, $^{c}$iknezevic@wisc.edu}}

\section{Introduction}

Thermal transport in semiconductors is governed by phonons, the quanta of lattice waves \cite{Ziman_Book,Balandin_JPCM_12}. At temperatures above several degrees Kelvin, phonons experience considerable multiphonon interactions, in addition to scattering with boundary roughness, interfaces, or atoms of different mass that stem from doping, alloying, or natural isotope variation. Therefore, phonon transport in nanostructures is typically in the quasiballistic or diffusive regimes, and is described well by the phonon Boltmann transport equation (PBTE). The PBTE can be solved via deterministic \cite{Broido_APL_2007,Aksamija_PRB_2010} or stochastic techniques \cite{Peterson_JHT_94, Mazumder_JHT_01, Lacroix_PRB_05} .

Phonon Monte Carlo (PMC) is an efficient stochastic technique for the solution to the PBTE   \cite{Peterson_JHT_94, Mazumder_JHT_01, Lacroix_PRB_05,Lemonnier_APL_06,
Baillis_JHT_08, Knezevic_PRB_12,Bera_JAP_2012,Knezevic_JAP_14,Peraud_AHRT_2014,Knezevic_APL_15,Ramiere_JPhysD_2016} which can incorporate real-space roughness and simulate nanostructures of experimentally relevant sizes. In PMC, a large ensemble of numerical phonons (typically of order 10$^5$ -- 10$^6$) is tracked over time as they fly freely and undergo scattering according to relevant scattering rates \cite{Glassbrenner_PR_1964}. Modern transport simulations based on the PBTE involve phonon-phonon scattering rates obtained from first principles \cite{Broido_APL_2007,Esfarjani_PRB_2011} and incorporate full phonon dispersions \cite{Knezevic_JAP_14}; the latter is very important in anisotropic systems, such as superlattices, nanowires, and nanoribbons \cite{Knezevic_JAP_14, Mei_JAP_2015}.

A number of random variables with generally complicated distribution functions underscore the behavior of the phonon ensemble. Examples of random variables include the phonon energy or momentum for a bulk phonon system in equilibrium, or the outgoing-momentum direction after a scattering event. The
PMC simulation hinges on the generation and use of \textit{random variates} -- specific values of the random variables that correspond to physical observables -- in a way that accurately and efficiently captures the appropriate distribution functions. Accurate and efficient generation of random variates that numerically represent nonuniform distribution functions is not a simple matter, yet most articles on PMC do not show much detail on this aspect.

In this chapter, we discuss numerical generation of random variates relevant for PMC, assuming that the uniformly distributed variates on the [0,1] interval are accessible. We discuss the relative merits of different approaches (direct inversion versus the rejection technique) from both theoretical and practical standpoints, which are sometimes at odds, and show several specific examples of nonuniform distributions relevant for phonon transport. We also identify common themes in phonon generation and scattering that are useful for reusing code in a simulation. We trust these examples will inform the reader about both the mechanics of random-variate generation and how to choose a good approach for whatever problem is at hand.

We review the two main methods for generating random variates in Sec. \ref{sec:random} and the PMC method in Sec. \ref{sec:intro}. Several applications are presented next: generating the attributes for phonons in equilibrium with full (Sec. \ref{sec:2Ddrawing}) and isotropic dispersions (Sec. \ref{sec:3Ddrawing}), randomizing outgoing momentum upon diffuse boundary scattering (Sec. \ref{sec:diffuseBC}), implementing contacts (Sec. \ref{sec:contacts}), and conserving energy in the simulations (Sec. \ref{sec:conservation}).

\section{Generating Random Variates}\label{sec:random}

Here, we present a quick overview of the methods used to generate random variates with a given probability distribution function (PDF). We assume that the computer used can generate random variates that are uniformly distributed over the interval $\left[0,1\right]$.

There are two techniques for generating nonuniform random variates from uniform random variates: the inversion and rejection methods, which we explain below. (For more details on random-variate generation, see a book such as \cite{Devroye_book}).

Generally speaking, the inversion method requires more analytical manipulation of the PDF than the rejection method. When the analytical manipulations are possible, the inversion method is usually the simpler of the two. While the inversion method can be performed numerically, the rejection technique is generally simpler to implement in cases when analytical inversion is not possible.

\subsection{Inversion Method}\label{inversion}

Consider a PDF $p\left(x\right)$. The first step in the inversion method is to integrate the PDF into the cumulative distribution function (CDF),

\begin{equation}
 F\left(x\right)=\int_{-\infty}^{x}dx^{\prime}f\left(x^{\prime}\right),
\end{equation}

\noindent
where $F\left(x\right)$ is the probability that a random variate will have a value less than or equal to $x$.

Next, we generate a random variate $r$, which is uniformly distributed in $\left[0,1\right]$. We solve $r=F\left(x\right)$ for $x$, i.e., we invert the CDF to get the quantile function $Q\left(r\right)=F^{-1}\left(r\right)$. Finally, we solve $x=Q(r)$ for $x$.The resulting $x$ is a random variate that follows our original PDF \cite{Devroye_book}. The technique can be generalized to PDFs with multiple variables, but we will only consider PDFs that  effectively only depend on a single variable.

For example, say we want to generate a random variate from the distribution given by Lambert's cosine law\footnote{This is also known as Knudsen's cosine law in the context of gas molecules scattering from surfaces.} in three dimensions (3D), $p\left(\theta,\phi\right) = c \cos{\theta}$ for spherical coordinates $\theta \in \left[0,\pi/2\right], \phi \in \left[0,2\pi\right)$ and a normalizing constant $c$ \cite{Modest_book}. Lambert's cosine law will prove important later (Secs. \ref{sec:diffuseBC} and \ref{sec:3Dboundary}). First, we must properly normalize $p\left(\theta,\phi\right)$:

\begin{equation}
 1 = \int_{0}^{\pi/2}\int_{0}^{2\pi} p\left(\theta^\prime,\phi^\prime\right) \sin{\theta^\prime} d\theta^\prime d\phi^\prime,
\end{equation}

\noindent
which yields $c=\pi^{-1}$. The CDF for $\theta$ is then

\begin{align}
    \begin{split}
        F\left(\theta\right) &= \int_{0}^{\theta}\int_{0}^{2\pi} p\left(\theta^\prime,\phi^\prime\right) \sin{\theta^\prime} d\theta^\prime d\phi^\prime \\
                             &= \sin^2 \theta.
    \end{split}
\end{align}

\noindent
Note that our CDF only depends on $\theta$. We can do this since $p\left(\theta,\phi\right)$ does not depend on $\phi$, but we keep the $\phi$ dependence to make clear that we still need to integrate over $\phi$. Finally, invert the CDF for a random variate $r_\theta$ that is uniformly distributed in $\left[0,1\right]$:

\begin{align}\label{equ:3Dlambert}
    \begin{split}
        F\left(\theta\right) &= r_\theta \,. \\
        \theta  &= \arcsin\left(\sqrt{r_{\theta}}\right).
    \end{split}
\end{align}

\noindent Using the same method, we can find the unsurprising result that $\phi=2 \pi r_\phi$, where $r_\phi$ is uniformly distributed in $\left[0,1\right]$.

In this example, both the integration and inversion can be done analytically, which is the exception rather than the rule. It is possible to do both the integration and inversion numerically, but this reduces the accuracy and computational efficiency of the method. We will see cases where both the integration and inversion are done numerically (Sec. \ref{sec:2Ddrawing}) and where the integration can be done analytically, but the inversion is done numerically (Sec. \ref{sec:diffuseBC}).

\subsection{Rejection Method}\label{sec:rejection}
In contrast with the inversion method, the rejection method does not require any integration or inversion steps. The rejection method requires that we calculate the PDF $p\left(x\right)$ for any $x$, there exists a bounding function $g\left(x\right)$ such that $\forall x, g\left(x\right) \geq p\left(x\right)$, and that we can generate random variates from a PDF that is proportional to $g\left(x\right)$. $p\left(x\right)$ and $g\left(x\right)$ do not have to be normalized; they must simply be proportional to probability distribution functions. The largest drawback of the rejection method is that, unlike the inversion method, the rejection method generally requires the computer to generate several random variates that are uniformly distributed in $\left[0,1\right]$. Choosing a $g\left(x\right)$ that closely resembles $p\left(x\right)$ will reduce the number of random variates that the computer will have to generate.

The rejection method to generate a random variate from $p\left(x\right)$ follows.

\begin{enumerate}
 \item Generate a random variate $x^\prime$ from the PDF that is proportional to $g\left(x\right)$.
 \item Generate a random variate $y$ that is uniformly distributed in $\left[0,g\left(x^\prime\right)\right]$
 \item If $y < p\left(x^\prime\right)$, then $x^\prime$ is the random variate generated from $p\left(x\right)$. Otherwise, return to step (1).
\end{enumerate}

\noindent The third step ensures that the probability of choosing $x^\prime$ is proportional to $p\left(x^\prime\right)$, which is all that we require of a method to generate random variates from a distribution.

\begin{figure}
	\includegraphics[width = \columnwidth]{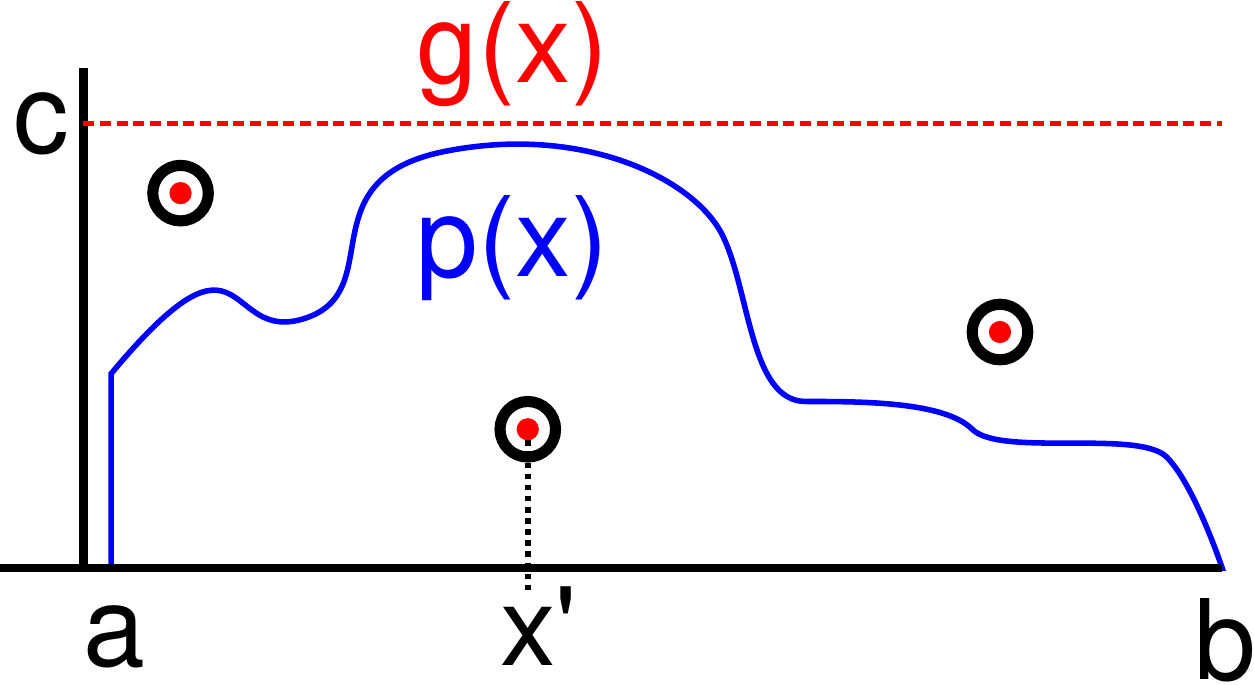}
	\caption{Illustration of the rejection method for a distribution $p\left(x\right)$ using a constant $g\left(x\right)$. For a constant $g\left(x\right)$, the rejection technique is the same as randomly throwing darts uniformly in the range $x \in \left[a,b\right], y \in \left[0,c\right]$. If the dart lands above the curve $p\left(x\right)$, then another dart is thrown. When a dart lands below the curve $p\left(x\right)$, then its $x$ value, $x^\prime$, is the random variate that is accepted. The figure depicts possible dart throws (marked with bullseyes). The throws above $p\left(x\right)$ are rejected, and $x^\prime$ is taken from the throw that lands below $p\left(x\right)$.}
	\label{fig:rejection}
\end{figure}

This procedure is easy to visualize if $p\left(x\right)$ is non-zero only on a finite interval $\left[a,b\right]$, and $p\left(x\right) \leq c$, where $a,b,c$ are constants (Fig. \ref{fig:rejection}). In this case, we can choose $g\left(x\right)=c$. Then the rejection method is equivalent to throwing a dart randomly and uniformly at the box defined by $x^\prime \in \left[a,b\right], y \in \left[0,c\right]$. If $y \leq p\left(x^\prime\right)$, i.e., the dart falls below the curve $p\left(x\right)$, then we take $x^\prime$ as our random variate and repeat the procedure otherwise. The larger the area between the line $y=c$ and the curve $p\left(x\right)$, the more dart throws will be required for a dart to land below the curve $p\left(x\right)$, which reduces the efficiency of the rejection method. For this reason, even if $p\left(x\right) \leq c$, it may be wise to use a $g\left(x\right)$ other than $g\left(x\right)=c$. We do this in Sec. \ref{sec:diffuseBC} when we consider Soffer's model for momentum-dependent boundary scattering \cite{Soffer_JAP_67}.

We note that the rejection technique can work even if $p\left(x\right)$ diverges, which is common in physics problems (e.g., Van Hove singularities). Take the case of a probability distribution that might arise from the Bose-Einstein distribution in two dimensions (2D) and in polar coordinates:

\begin{equation}
 p\left(r, \theta \right) = \frac{1}{e^{r} - 1}.
\end{equation}

\noindent Note that $p\left(r, \theta\right)$ is not normalized and diverges at $r=0$. Also note that $p\left(r, \theta\right) \leq r^{-1}$, so we choose $g\left(r, \theta\right)=r^{-1}$. Finally, suppose that we are considering the 2D domain $\theta \in \left[0,2\pi\right), r \in \left[0,R\right]$, where $R$ is a constant. On that domain, $g\left(r, \theta\right)$ and $p\left(r, \theta\right)$ are normalizable because the singularity at $r=0$ is integrable: $\int^{2\pi}_0 \int^R_0 g\left(r, \theta\right) r dr d\theta = 2 \pi R$, which is finite. Because the integrand $g\left(r, \theta\right) r$ is constant, the random variate $r^\prime$ is equally likely to take any value in $\left[0,R\right]$. So, it is quite simple to generate the random variates $r^\prime$ and $\theta^\prime$: $r^\prime$ is uniformly distributed in $\left[0,R\right]$, and $\theta^\prime$ is uniformly distributed in $\left[0, 2\pi\right)$.

Putting everything together, the rejection technique in this example works as follows:

\begin{enumerate}
 \item Generate random variates $r^\prime$ and $\theta^\prime$ that are uniformly distributed in $\left[0,R\right]$ and $\left[0,2 \pi\right)$, respectively.
 \item Generate a random variate $y$ that is uniformly distributed in $\left[0,\left(r^\prime\right)^{-1}\right]$
 \item If $y < p\left(r^\prime, \theta^\prime\right) = \left(e^{r^\prime} - 1\right)^{-1}$, then use $r^\prime$ and $\theta^\prime$ as your random variates. Otherwise, return to step (1).
\end{enumerate}

\section{Overview of Phonon Monte Carlo}\label{sec:intro}

Phonons are the main carriers of heat in semiconductor materials \cite{Balandin_JPCM_12}. Phonons can be treated as semiclassical particles on the spatial scales longer than the phonon coherence length and time scales longer than the phonon relaxation time. Phonon transport under these conditions is captured via the phonon Boltzmann transport equation (PBTE) \cite{Ziman_Book}:
\begin{equation} \label{equ:PBTE}
\frac{\partial n_\mathrm{b}(\bm{r},\bm{q},t)}{\partial t} + \bm{v}_{\mathrm{b},\bm{q}}\cdot\nabla_{\bm{r}}n_\mathrm{b}(\bm{r},\bm{q},t)=\left.\frac{\partial n_\mathrm{b}(\bm{r},\bm{q},t)}{\partial t}\right|_{\mathrm{scat}}.
\end{equation}
$n_\mathrm{b}(\bm{r},\bm{q},t)$ is the time-dependent distribution of phonons with respect to position $\bm{r}$ and the phonon wave vector $\bm{q}$ for phonon branch b. \(\bm{v}_{\mathrm{b},\bm{q}}=\nabla_{\bm{q}}\omega_{\mathrm{b},\bm{q}}\) is the phonon group velocity, where $\omega_{\mathrm{b},\bm{q}}$ is the phonon angular frequency in branch b at wave vector $\bm{q}$. In equilibrium, the average occupancy of a phonon state with energy $\hbar\omega$ at absolute temperature $T$ is given by the Bose-Einstein distribution function
\begin{equation}\label{equ:BE}
\langle n_\mathrm{BE}(\omega,T)\rangle=\frac{1}{e^{\frac{\hbar\omega}{k_\mathrm{B}T}}-1},
\end{equation}
where $k_B$ is the Boltzmann constant. When addressing out-of-equilibrium phonon transport, we assume local equilibrium and employ the concept of a local and instantaneous temperature, $T(\bm{r},t)$, then calculate the expectation number of phonons accordingly.

\begin{figure}
	\includegraphics[width = 0.9\columnwidth]{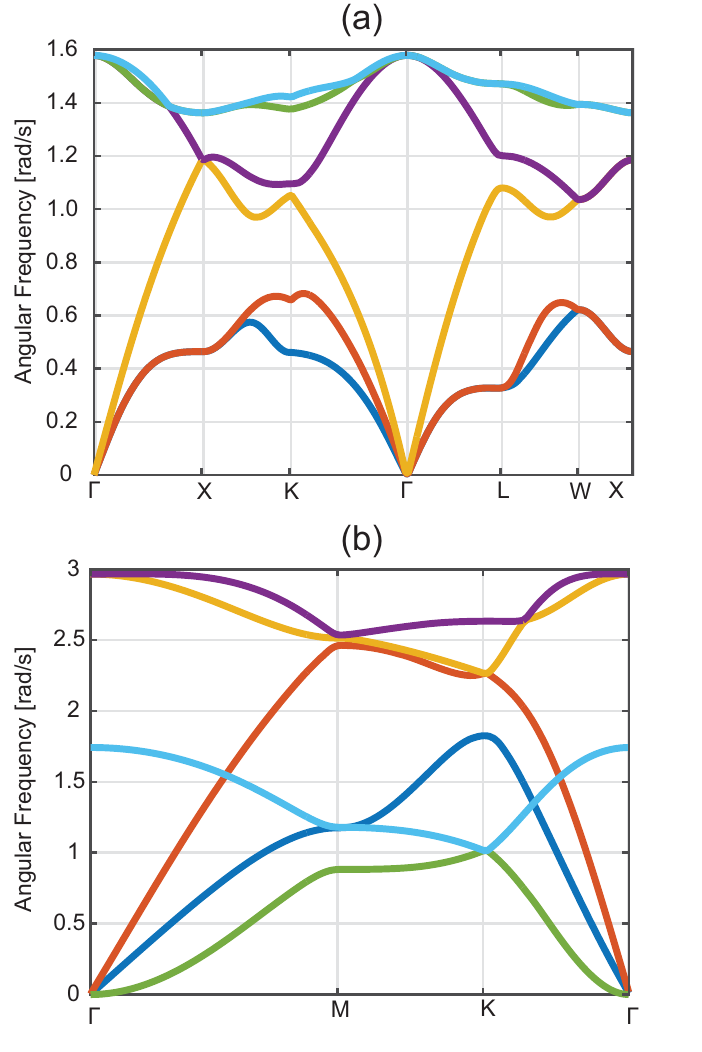}
	\centering
	\caption{Phonon dispersions for (a) 3D silicon and (b) 2D graphene along high-symmetry directions.}
	\label{fig:disp}
\end{figure}

To obtain the phonon group velocity $\bm{v}_{\mathrm{b},\bm{q}}$, we require the phonon dispersion relations -- the relationships between the phonon angular frequency $\omega_{\mathrm{b},\bm{q}}$ and the phonon wave vector $\bm{q}$. Figure~\ref{fig:disp} shows the full phonon dispersions along the high-symmetry directions in 3D silicon and 2D graphene. In acoustic branches, the phonon dispersion is relatively isotropic, therefore an analytical isotropic phonon dispersion ($\omega_{\mathrm{b},\bm{q}}$ depends only on the magnitude of $\bm{q}$) is often adopted, as it enables a direct and simple evaluation of the phonon group velocity from the random phonon wave vector obtained after scattering (more on this in Sec.~\ref{sec:random}), which speeds up the simulation. A quadratic isotropic dispersion\footnote{$\omega_\mathrm{b}(\bm{q})=v_{s,\mathrm{b}}q+c_\mathrm{b}q^2$, where $q=|\bm{q}|\in[0,q_\mathrm{max}]$ is the allowed wave number, $v_{s,\mathrm{b}}$ is the mode-dependent sound velocity, and $c_\mathrm{b}$ is quadratic coefficient fitted from the full dispersion. The maximum allowed wave number $q_\mathrm{max}$ is calculated from $\frac{2\pi}{a_0}$, where $a_0$ is the lattice constant of the material.} has been shown to be very accurate in simulating the thermal properties of silicon \cite{Mazumder_JHT_01,Lemonnier_APL_06,Knezevic_PRB_12}.

However, when the simulated structure is very small, such as in the case of thin nanowires or graphene nanoribbons (GNRs), anisotropy is prominent and adopting the full dispersion is necessary at the cost of slowing down the simulation. In the following sections, the isotropic approximation and the full dispersion relations in the context of PMC will be demonstrated on the examples of 3D silicon and 2D graphene, respectively.

\begin{figure}
	\includegraphics[width = \columnwidth]{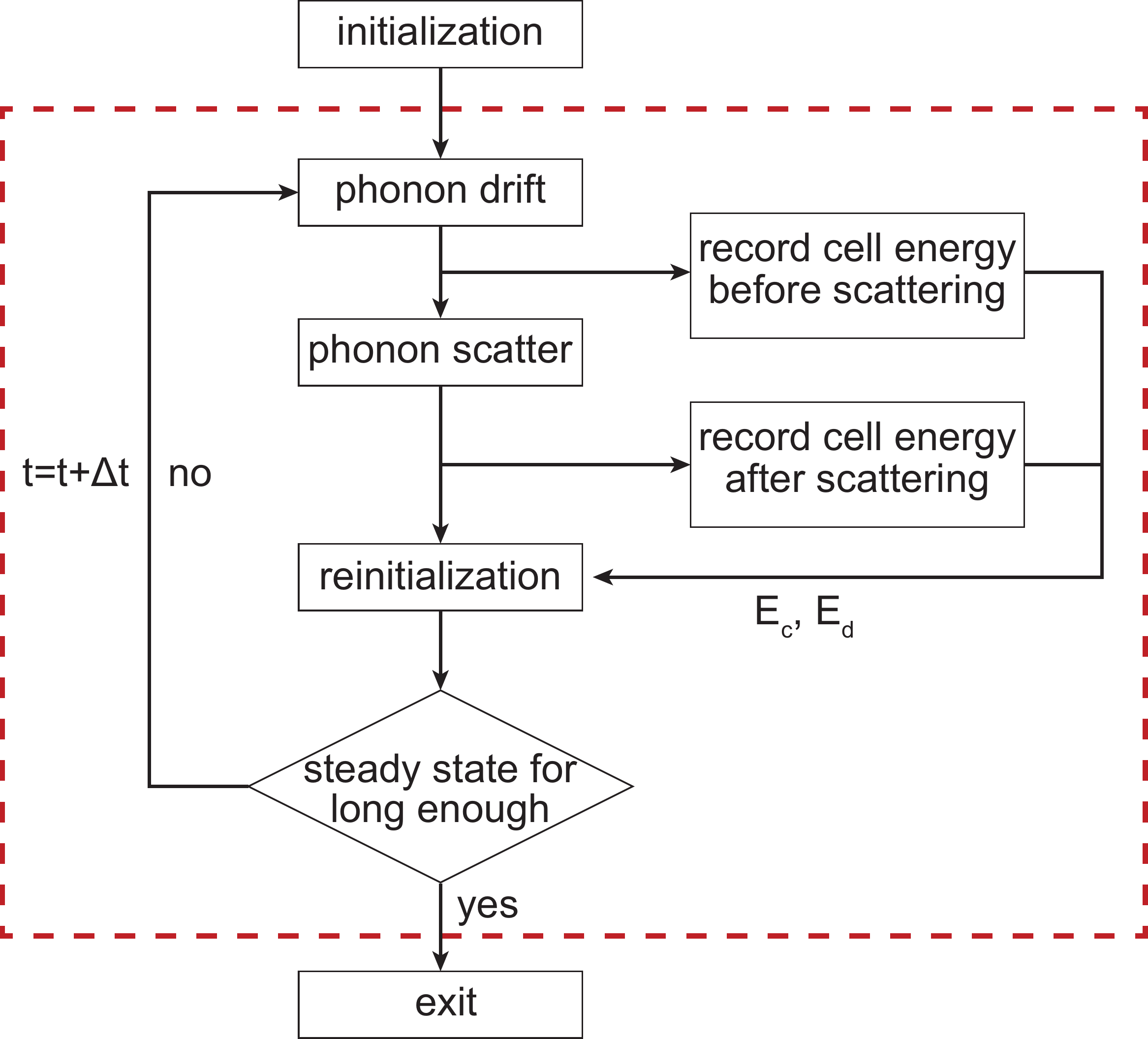}
	\caption{Flowchart of a PMC simulation. The dashed box encloses the transport kernel.}
	\label{fig:flowchart}
\end{figure}

\begin{figure}
	\includegraphics[width = \columnwidth]{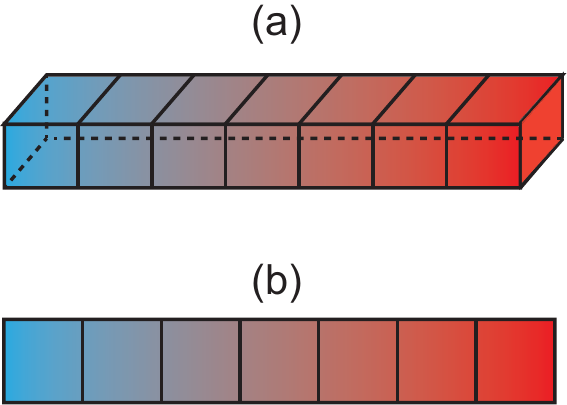}
	\caption{A typical simulation domain for a (a) 3D and (b) 2D PMC simulation. The color represents a typical steady-state temperature profile in each structure when connected to cold (blue, left end) and hot (red, right end) reservoirs.}
	\label{fig:domain}
\end{figure}

Phonon Monte Carlo is a widely used stochastic technique for solving the PBTE and obtaining thermal-transport properties of semiconductor materials \cite{Mazumder_JHT_01,Lemonnier_APL_06}. PMC tracks phonon transport in real space, thus allowing easy implementation of nontrivial geometries such as rough boundaries and real-space edge structures \cite{Knezevic_PRB_12,Knezevic_JAP_14,Knezevic_APL_15}. Figure~\ref{fig:flowchart} shows the flowchart of a typical PMC simulation. The simulation domain is a wire with a square cross-section (3D) or a rectangle (2D) divided into $N_\mathrm{c}$ cells of equal length along the heat-transport direction, as shown in Fig. \ref{fig:domain}. The two end cells are connected to heat reservoirs fixed at slightly different temperatures ($T_h$ and $T_c$). For structures much longer than the phonon mean free path (i.e., in the diffusive-transport limit) and in a steady state, a linear temperature profile between $T_h$ and $T_c$ will naturally develop inside the simulation domain (Fig. \ref{fig:domain}). Therefore, we can achieve a steady state in the simulation faster if we initialize the cells according to a linear temperature profile, with the $i$th cell temperature set to \(T_i=T_h-\frac{i-1}{N_\mathrm{c}-1}(T_h-T_c)\).

The total energy associated with the $i$th cell is then
\begin{equation}\label{equ:eoriginal}
\mathcal{E}_i=\Omega_i\sum\limits_{\mathrm{b}} \int D_\mathrm{b}(\omega)\langle n_\mathrm{BE}(\omega,T_i)\rangle\hbar\omega~ d\omega,
\end{equation}
where $\Omega_i$ is the volume (3D) or area (2D) of the $i$th cell in real space, $D_\mathrm{b}(\omega)$ is the phonon density of states (PDOS) in the material associated with branch b and energy level $\hbar\omega$, and the sum is over all three acoustic branches. The expectation number of phonons in the $i$th cell is then
\begin{equation}\label{equ:numberDOS}
\mathcal{N}_{i,\mathrm{exp}}=\Omega_i\sum\limits_{\mathrm{b}} \int D_\mathrm{b}(\omega)\langle n_\mathrm{BE}(\omega,T_i)\rangle~d\omega.
\end{equation}

\noindent In simulations where the sample size is large (on the order of microns) or the temperature is not very low (a few hundred Kelvin), the expectation number of phonons $\mathcal{N}_{i,\mathrm{exp}}$ can be very high ($10^7-10^9$), and it is computationally expensive to keep track of this many particles in the simulation. Instead, a weighting factor $W$ is often introduced \cite{Mazumder_JHT_01,Lemonnier_APL_06} to reduce the number of simulation particles to a tolerable range (typically \(10^5-10^6\)).

In 2D materials, like graphene, the number of flexural out-of-plane (ZA) phonons is overwhelmingly larger than that of transverse acoustic (TA) or  longitudinal acoustic (LA) phonons, owing to the shape of the dispersion curves; therefore, a branch-dependent weighting factor $W_{\mathrm{b}}$ should be used. With the weighting factor taken into consideration, the total energy of the simulation particles in a cell becomes
\begin{equation}\label{equ:enrcontinue}
E_i=\Omega_i\sum\limits_{\mathrm{b}} \int D_\mathrm{b}(\omega)\frac{\langle n_\mathrm{BE}(\omega,T_i)\rangle}{W_\mathrm{b}}\hbar\omega~ d\omega,
\end{equation}

\noindent During the initialization, we keep generating phonons following a desired distribution according to each cell's temperature and add them to random positions inside the cell until the cell has the desired energy, as expressed by Eq.~(\ref{equ:enrcontinue}). After initialization, we enter the transport kernel (enclosed in the dashed box in Fig.~\ref{fig:flowchart}), where time is discretized in steps of $\Delta t$. Each phonon is allowed to drift according to the group velocity obtained through the dispersion relation. A random number is drawn to decide whether the phonon will be scattered during its drift; the possibility of being scattered is captured through a phonon relaxation time. We record the heat flux along the wire (3D)/ribbon (2D) to monitor whether a steady state has been reached (i.e., whether the flux has become constant) and use ensemble averages to compute the thermal properties of the material. Special measures (e.g., reinitialization) are needed to make sure the energy in each cell is properly conserved without violating the distribution (Sec. \ref{sec:conservation}). More details about the full-dispersion PMC simulation can be found in \cite{Knezevic_JAP_14}.

\section{Generating Phonon Attributes in PMC}\label{sec:generating}

\subsection{Thermal Phonons with Full Dispersion in 2D}\label{sec:2Ddrawing}

With the basic knowledge of the methods to generate random variates following certain distributions, this section gives examples of using these methods to randomly draw a phonon from the equilibrium distribution, following the full dispersion relation in 2D graphene.

Here, we will present both the inversion technique with numerical integration and inversion and the rejection technique; the former is commonly used for thermal phonons with isotropic dispersion in 3D \cite{Mazumder_JHT_01, Lacroix_PRB_05, Knezevic_PRB_12}, as is the rejection technique  \cite{Peterson_JHT_94,Knezevic_APL_15}.  The inversion technique is used to choose the phonon frequency and branch, and the rejection technique is used to choose a wave vector that matches the chosen frequency and branch. We note that we could use the rejection technique to choose the branch and wave vector directly (the method would be similar to the example at the end of Sec.~\ref{sec:rejection}), but we choose a hybrid approach to allow for code reuse: for internal scattering, we already need the code that can generate phonons of a specific frequency. So, we use the inversion technique to choose the phonon frequency and then use the preexisting code to draw a corresponding wave vector.

As introduced in Sec.~\ref{sec:intro}, we have assigned a temperature to the cell that we are generating a phonon in. The first step in generating the phonon is to find an angular frequency which follows the Bose-Einstein distribution according to the temperature. To do that, we use the CDF of $\omega$ at the given temperature $T$ with the help of DOS:
\begin{equation}\label{equ:cdfcont}
F(\omega,T)=\frac{\sum_{\mathrm{b}}\int_{0}^{\omega} d\omega\langle n_\mathrm{BE}(\omega,T)\rangle D_\mathrm{b}(\omega)/W_\mathrm{b}}{\sum_{\mathrm{b}}\int_{0}^{\omega_\mathrm{max}}d\omega \langle n_\mathrm{BE}(\omega,T)\rangle D_\mathrm{b}(\omega)/W_\mathrm{b}}.
\end{equation}
With full dispersion, we do not have an analytical expression for $D_\mathrm{b}(\omega)$. Therefore, we divide the range of $[0,\omega_\mathrm{max}]$ into $N_\mathrm{int}$ equal energy bins where \(\Delta\omega=\lceil\frac{\omega_\mathrm{max}}{N_\mathrm{int}}\rceil\) is the interval length and $\omega_{\mathrm{c},i}=\frac{2i-1}{2}\Delta\omega$ is the central frequency of the $i$th interval. We can obtain $D_\mathrm{b}(\omega_{\mathrm{c},i})$ for $i=1,2,\ldots,N_\mathrm{int}$ and evaluate the discrete CDF as
\begin{equation}\label{equ:pomega}
F_i(T)=\frac{\sum\limits_{\mathrm{b}}\sum\limits_{j=1}^{i} \langle n_\mathrm{BE}(\omega_{\mathrm{c},j},T)\rangle D_\mathrm{b}(\omega_{\mathrm{c},j})/W_\mathrm{b}}{\sum\limits_{\mathrm{b}}\sum\limits_{j=1}^{N_\mathrm{int}} \langle n_\mathrm{BE}(\omega_{\mathrm{c},j},T)\rangle D_\mathrm{b}(\omega_{\mathrm{c},j})/W_\mathrm{b}}.
\end{equation}
\begin{figure}
	\includegraphics[width = \columnwidth]{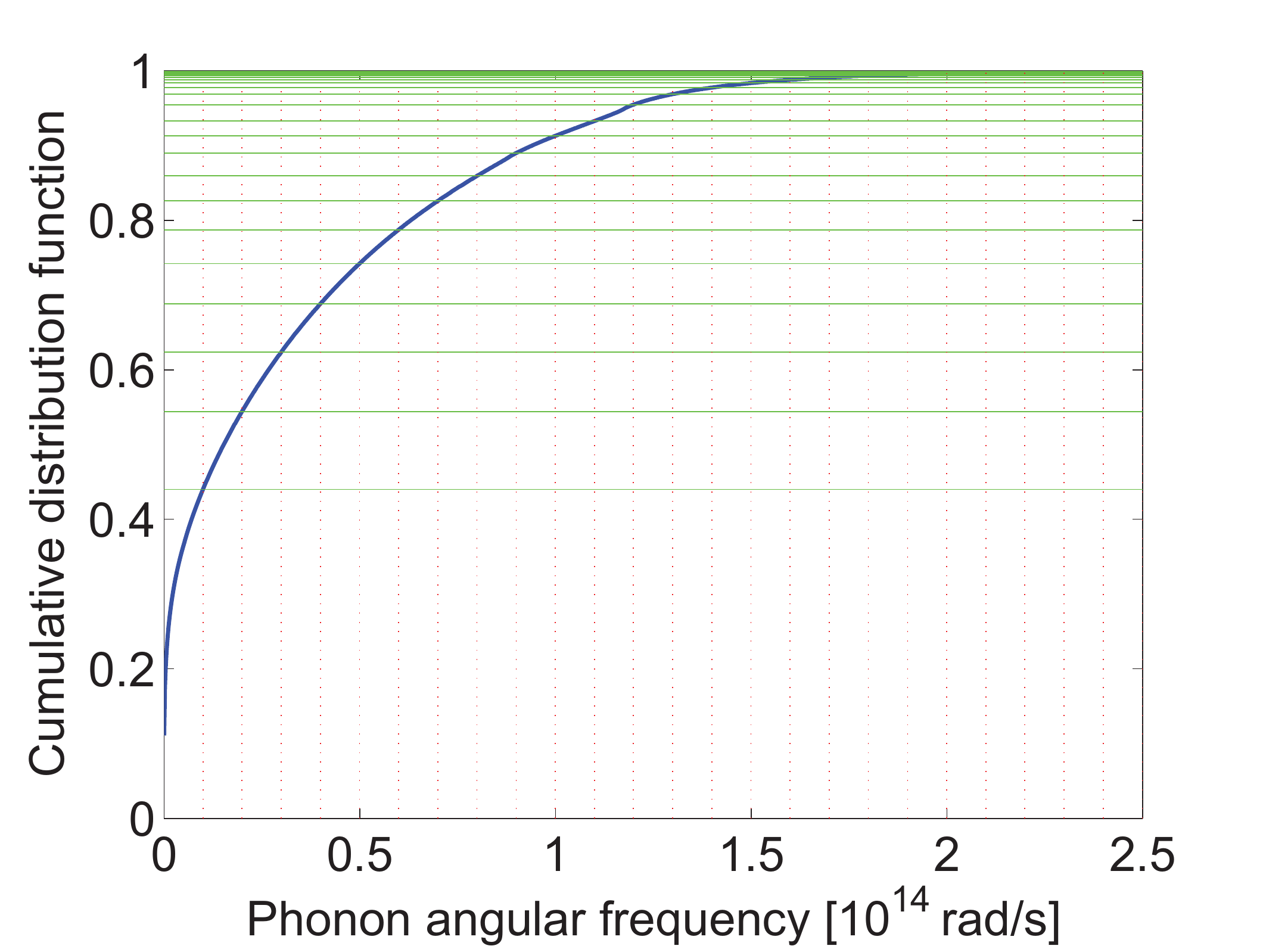}
	\caption{Cumulative distribution function of angular phonon frequency at 300 K. The frequency range $\omega\in[0,2.5\times 10^{14}]$~rad/s was divided into $N_\mathrm{int}=2500$ equal intervals in the numerical calculation. Reprinted with permission from S. Mei, L. N. Maurer, Z. Aksamija, and I. Knezevic, J. Appl. Phys. 116, 164307 (2014).  Copyright 2014, American Institute of Physics.}
	\label{fig:CDF}
\end{figure}
To complete the table, set \(F_\mathrm{0}(T)=0\), meaning that all phonons must have positive energy. The numerically evaluated CDF for phonons at 300 K is shown in Fig.~\ref{fig:CDF}.

The next step is to numerically invert the the table: we draw a random number \(r_1\) and look for the interval $i$ satisfying \(F_{i-1}<r_1<F_i\). We decide the frequency of this phonon falls in the $i$th interval and the actual frequency is determined with another random number \(r_2\),  \(\omega=\omega_{\mathrm{c},i}+(2r_2-1)\frac{\Delta\omega}{2}\).

When the $\omega$ is chosen, phonon branch b can be chosen in a similar fashion. Use index 1, 2, and 3 to represent the TA, LA, and ZA branches, respectively. Since there are only 3 branches, we can enumerate the CDFs as
\begin{subequations}\label{equ:branch_choice}
	\begin{eqnarray}
	F_1(\omega) &=& \frac{D_1(\omega)/W_1}{\sum_\mathrm{b'}D_\mathrm{b'}(\omega)/W_\mathrm{b'}},\\
	F_2(\omega) &=& \frac{D_1(\omega)/W_1+D_2(\omega)/W_2}{\sum_\mathrm{b'}D_\mathrm{b'}(\omega)/W_\mathrm{b'}},\\
	F_3(\omega) &=& 1,
	\end{eqnarray}
\end{subequations}
and a third random number $r_3$ is used to choose the phonon branch b.

\begin{figure}
	\includegraphics[width = \columnwidth]{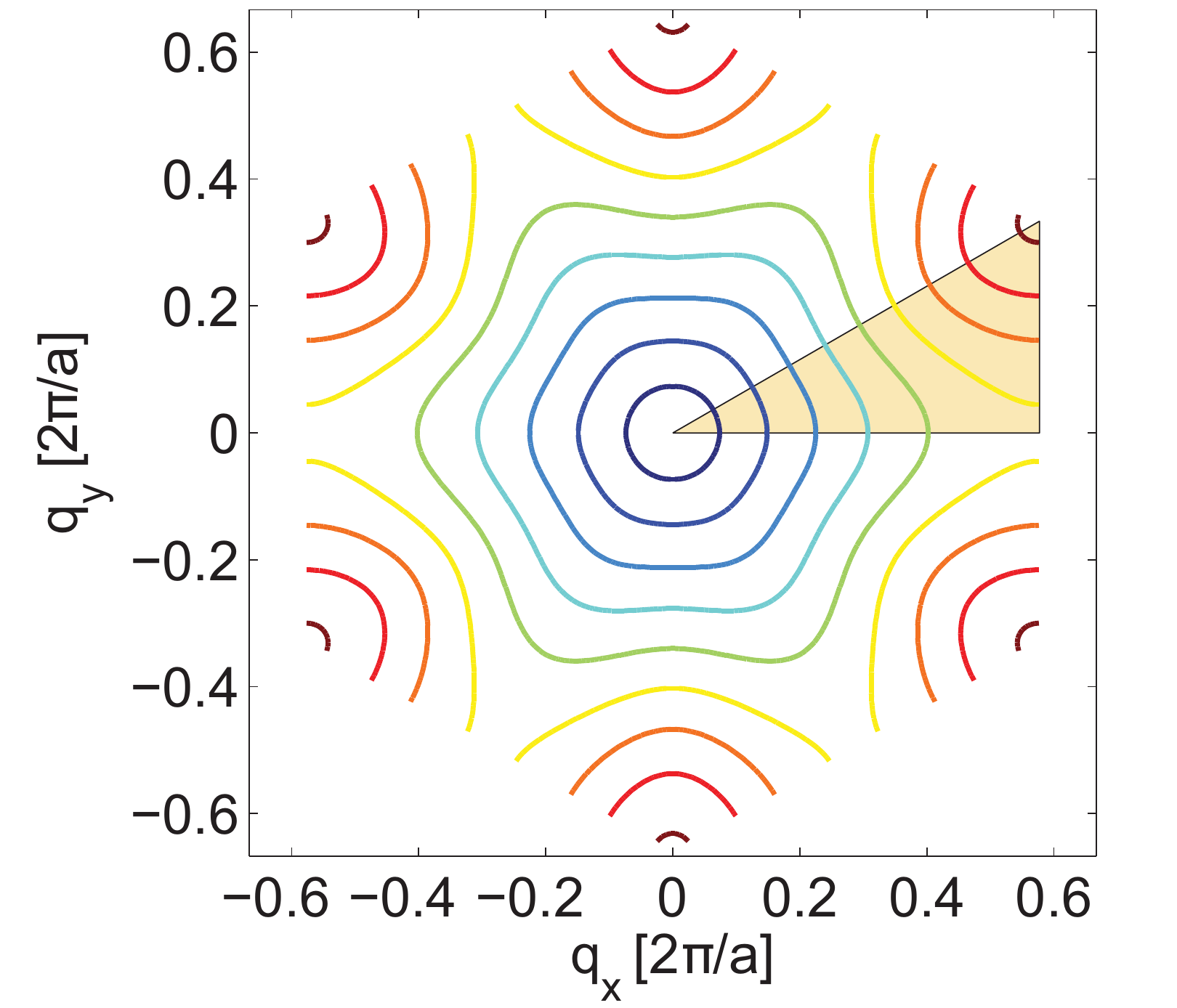}
	\caption{Isoenergy curves in the first Brillouin zone for TA-branch phonons. Adjacent curves are offset by $2\times 10^{13}$ rad/s and the shaded region in the black triangle is the irreducible wedge in the first Brillouin zone. Reprinted with permission from S. Mei, L. N. Maurer, Z. Aksamija, and I. Knezevic, J. Appl. Phys. 116, 164307 (2014).  Copyright 2014, American Institute of Physics.}
	\label{fig:eqpot}
\end{figure}

The next step is generating the phonon wave vector, $\bm{q}$, for the $\omega$ and b we already found. The distribution is 2D and complex, so it is hard to calculate a CDF. As a result, we employ the rejection technique. Figure \ref{fig:eqpot} shows the isoenergy curves for the TA branch in the first Brillouin zone (1BZ), adjacent curves differing by $2\times 10^{13}$ rad/s.
Because the isoenergy curves are close to circles, it is convenient to use the polar coordinate. Further, because of the symmetry, we can generate $\bm{q}$ in the shaded area and simply map it to the whole 1BZ. (More details can be found in \cite{Knezevic_JAP_14}.) We use the rejection technique to choose an angle $\theta\in[0,\frac{\pi}{6}]$ and use a lookup table to get the corresponding $|\bm{q}|$.
The probability of a phonon with frequency $\omega$ in branch b having angle $\theta$ is represented by
\begin{equation}\label{equ:p}
p(\omega,\theta)\propto\frac{arc(\theta-\delta\theta,\theta+\delta\theta)}{\left|\bm{v}_\mathrm{g}(\omega,\theta)\right|},
\end{equation}
where \(arc(\theta-\delta\theta,\theta+\delta\theta)\) is the arc length on the isoenergy curve between \((\theta-\delta\theta,\theta+\delta\theta)\) and  \(\left|\bm{v}_\mathrm{g}(\omega,\theta)\right|\) is the magnitude of group velocity. When $\delta\theta$ is small, it is acceptable to assume the group velocity is constant along the arc. Since we do not have an analytical expression for the probability, we make a rejection table of $N_\mathrm{a}$ equally separated points between $[0,\frac{\pi}{6}]$ where $\Delta\theta=\frac{\theta_\mathrm{max}}{N_\mathrm{a}}$ is the spacing and $\theta_{c,i}=(2i-1)\frac{\Delta\theta}{2}$ is the central frequency in the $i$th interval. Note that energy is also discrete; we evaluate Eq.~(\ref{equ:p}) only at ($\omega_{\mathrm{c},i},\theta_{\mathrm{c},j}$) and obtain an \(N_\mathrm{int}\times N_\mathrm{a}\) interpolation table. For any $0<\omega<\omega_\mathrm{max}$ and $0<\theta<\theta_\mathrm{max}$ we can get the probability of having a phonon from interpolation. $N_\mathrm{a}=100$ is enough for accurate interpolation. Upon getting the energy $\omega$, we get the $1\times N_\mathrm{a}$ angle-distribution table $f_\omega(\theta_{\mathrm{c},i})$ and record $f_{\omega,\mathrm{max}}=\max_{i=0}^{N_\mathrm{a}}\left(f_\omega(\theta_{\mathrm{c},i})\right)$. A pair of uniformly distributed random numbers $(x,y)$ is drawn. Therefore, our angle in consideration is $\alpha=\frac{\pi}{6}\cdot x$. Then interpolate the distribution table to obtain $f_\omega(\alpha)$. If $f_{\omega,\mathrm{max}}\cdot y\leq f_\omega(\alpha)$, the angle $\alpha$ is accepted, and we can proceed to look up $|\bm{q}|(\omega,\alpha)$ by interpolation. The final chosen wave vector is then $(|\bm{q}|(\omega,\alpha)\cos\alpha, |\bm{q}|(\omega,\alpha)\sin\alpha)$. If  $f_{\omega,\mathrm{max}}\cdot y>f_\omega(\alpha)$, the angle is rejected. We keep generating $(x,y)$ pairs until an angle is accepted, which typically occurs within two iterations with the simple $f_{\omega,\mathrm{max}}$ bound.

\subsection{Thermal Phonons with an Isotropic Dispersion in 3D}\label{sec:3Ddrawing}

As we mentioned in the previous section, it is common to draw thermal phonons  with isotropic dispersion relations in 3D using the inversion technique with numerical integration and inversion \cite{Mazumder_JHT_01, Lacroix_PRB_05, Knezevic_PRB_12}, although the rejection method is also sometimes used \cite{Peterson_JHT_94,Knezevic_APL_15}. We believe that the rejection method is better suited to the task than the inversion method because the numerical integration and inversion can lead to a loss off accuracy or decreased computational performance \cite{Mazumder_JHT_01}. Additionally, the rejection method does not require the density of states, which can be difficult to calculate. We will consider two branches, a TA branch and an LA branch, but the method can easily be generalized to more branches. We will also assume $q_\mathrm{max} \ge \left|\bm{q}\right|$ is an upper bound on the wave-vector magnitude. The rejection method works as follows.


At temperature $T$, the number of phonons from branch b in each infinitesimal unit of reciprocal space is

\begin{equation}
 n_\mathrm{b} \left(\bm{q}, T\right) = \frac{d^3\bm{q}}{\left(2 \pi\right)^3} \langle n_\mathrm{BE}(\omega_{\mathrm{b},\bm{q}},T)\rangle.
\end{equation}

\noindent For an isotropic dispersion relation, $n_\mathrm{b}$ can be reduced to a function of $q$ alone by integrating over the polar and azimuthal angles $\theta$ and $\phi$

\begin{align}
 \begin{split}
  n_\mathrm{iso,b} \left(q,T\right) &=  \int_0^{2\pi} \int_0^{\pi} \frac{q^2\sin{\theta} dq d\theta d\phi}{\left(2 \pi\right)^3} \langle n_\mathrm{BE}(\omega_{\mathrm{b},\bm{q}},T)\rangle \\
  &= \frac{q^2}{2 \pi} \langle n_\mathrm{BE}(\omega_{\mathrm{b},q},T)\rangle dq.
 \end{split}
\end{align}

\noindent Although $\langle n_\mathrm{BE}\rangle$ diverges at $q=0$, $\lim_{q \to 0} n_\mathrm{b} \left(\bm{q}\right)=0$ because of the $q^2$ term. Thus, the maximum value of $n_\mathrm{iso, b} \left(q,T\right)$ is finite, and we can use the bounding function $g\left(q\right)=c$ as described in Sec. \ref{sec:rejection}, where $c$ is any number greater than the maximum value of $n_\mathrm{iso,TA} \left(q,T\right) + n_\mathrm{iso,LA} \left(q,T\right)$. The maximum value of $n_\mathrm{b} \left(q,T\right)$ is temperature dependent, but instead of finding a new $c$ whenever the temperature changes, simply find a $c$ that works for a temperature higher than any conceivable temperature in your simulation. Once $c$ is found, the rejection method follows the familiar pattern:

\begin{enumerate}
 \item Generate a random variate $q^\prime$ that is uniformly distributed in $\left[0,q_\mathrm{max}\right]$
 \item Generate a random variate $y$ that is uniformly distributed in $\left[0, c\right]$
 \item If $y < n_\mathrm{iso,TA} \left(q,T\right)$, then generate a TA phonon with wave number $q^\prime$. If $n_\mathrm{iso,TA} \left(q^\prime,T\right) < y < n_\mathrm{iso,TA} \left(q^\prime,T\right) + n_\mathrm{iso,LA} \left(q^\prime,T\right)$, then generate an LA phonon with wave number $q^\prime$. Otherwise, return to step (1).
\end{enumerate}

\noindent Choosing between different branches in the last step is similar to the procedure used for choosing a branch in Eq.~(\ref{equ:branch_choice}).

Once we know the wave number $q$ of the new phonon, we need to choose a direction. In equilibrium, all directions are equally likely [$p\left(\theta, \phi\right) = d$, where $d$ is a constant], and we can use the inversion technique to choose a direction. First, we find $d$

\begin{align}
 \begin{split}
  1 &= \int_{0}^{\pi}\int_{0}^{2\pi} p\left(\theta^\prime,\phi^\prime\right) \sin{\theta^\prime} d\theta^\prime d\phi^\prime \\
  d &= \left(4 \pi \right)^{-1}.
 \end{split}
\end{align}

\noindent The CDF is\footnote{In the CDF, $\left(1-\cos{\theta}\right)/2$ can be rewritten as $\sin^2 (\theta /2)$, but making that change has no advantage in implementing the code, although it does make the equation look more like the results from Lambert's cosine law (Eq. (\ref{equ:3Dlambert})).}

\begin{align}
 \begin{split}
  P\left( \theta \right) &= \int_{0}^{\theta}\int_{0}^{2\pi} p\left(\theta^\prime,\phi^\prime\right) \sin{\theta^\prime} d\theta^\prime d\phi^\prime \\
  &= \frac{1-\cos{\theta}}{2}.
 \end{split}
\end{align}

\noindent Inverting for a random variate $r_\theta$ that is uniformly distributed in $\left[0,1\right]$

\begin{equation}
 \theta = \arccos{\left(1-2 r_\theta\right)}.
\end{equation}

$\phi$ is uniformly distributed in $\left[0,2 \pi\right]$.

All together

\begin{equation}
\bm{q}=q^\prime \left(\begin{array}{c}
\sin\left(\theta\right)\cos\left(\phi\right)\\
\sin\left(\theta\right)\sin\left(\phi\right)\\
\cos\left(\theta\right)
\end{array}\right)
\end{equation}

\section{Diffuse Boundary Scattering}\label{sec:diffuseBC}

Nanostructure surfaces can have a large impact on thermal conductivity through increased boundary-surface scattering \cite{Yang_NLet_12}, so implementing real-space phonon--surface scattering is important for PMC simulations. A common class of phonon--surface scattering models rely on diffuse scattering and a specularity parameter: a phonon that strikes a surface either reflects specularly or is scattered into a new phonon of the same frequency and branch,  but with a random wave vector. The probability of specular scattering is controlled by a specularity parameter, which is either constant \cite{Ziman_PRS_53} or momentum dependent \cite{Ziman_PRS_55, Ziman_Book, Soffer_JAP_67}. Specularity-parameter models fail when the surfaces are very rough, but appear to work well for small roughnesses \cite{Knezevic_APL_15}. Additionally, real phonons of one branch can be scattered into phonons of a different branch at a surface \cite{Northrop_PRL_84}. This process, referred to as mode conversion, is often ignored, but can be included in phonon Monte Carlo simulations \cite{Antonsen_PRA_92}.

\begin{figure}
	\includegraphics[width = \columnwidth]{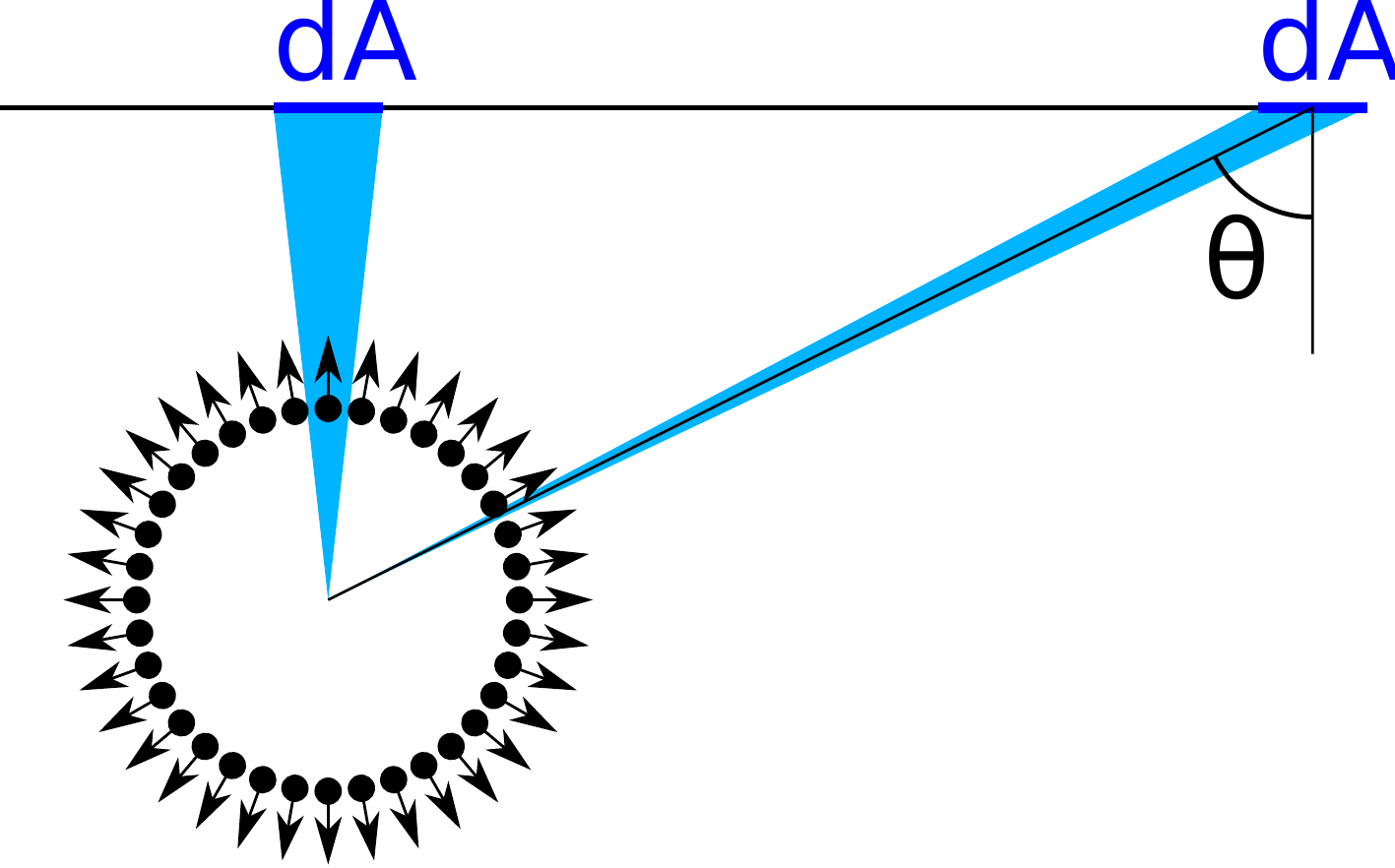}
	\caption{Illustration of Lambert's cosine law. An equilibrium ensemble of phonons is depicted as a group of particles with uniform angular distribution (lower left). The number of phonons striking a small patch of surface $dA$ at an angle $\theta$ (measured from the surface normal) is proportional to the angle subtended by the shaded wedges. This angle is proportional to $\cos \theta$ (if $dA$ is small), which is the basis for Lambert's cosine law.}
	\label{fig:Lambert}
\end{figure}

All phonon--surface scattering models must respect detailed balance in equilibrium: the number of phonons scattered into any solid angle must be equal to the number of phonons incident on the surface from the same solid angle. Consider a flat surface with a surface normal vector $\hat{n}$. In this section, $\theta$ will refer to the angle between a phonon's wave vector $\bm{q}$ and the surface's unit normal vector. (We will assume that the dispersion relation is isotropic, so that $\bm{q}$ is in the same direction as the velocity.) Phonons with a larger $\bm{q} \cdot \hat{n} = \left| \bm{q} \right| \cos{\theta}$ will strike the surface more frequently, and in equilibrium this leads to Lambert's cosine law: the number of phonons incident on a surface at an angle $\theta$ must also be proportional to $\cos{\theta}$ (Fig. \ref{fig:Lambert}). Because of detailed balance, the number of phonons being scattered into an angle $\theta$ is proportional to $\cos{\theta}$.

For specular scattering, detailed balance in equilibrium holds automatically. For totally diffuse scattering, the inversion method can beused to generate random wave vectors that satisfy Lambert's cosine law. Lambert's cosine law in 3D was an example in Sec. \ref{inversion} with the result given in Eq. (\ref{equ:3Dlambert}). In 2D, the procedure is very similar except that we work in polar rather than spherical coordinates. We use the distribution $p\left( \theta \right) = c \cos{\theta}$, where $c$ is a normalizing constant. The normalization condition is now

\begin{equation}
 1 = \int_{0}^{\pi/2} p\left(\theta^\prime\right) d\theta^\prime,
\end{equation}

\noindent
which yields $c=1$. The CDF is

\begin{align}
    \begin{split}
        F\left(\theta\right) &= \int_{0}^{\theta} p\left(\theta^\prime\right) d\theta^\prime \\
                             &= \sin{\theta}.
    \end{split}
\end{align}

\noindent Finally, we invert the CDF for a random variate $r_\theta$ that is uniformly distributed in $\left[0,1\right]$:

\begin{align}\label{equ:2Dlambert}
    \begin{split}
        F\left(\theta\right) &= r_\theta , \\
        \theta  &= \arcsin{\theta},
    \end{split}
\end{align}

\noindent
which resembles the 3D result.

For the constant-specularity-parameter model in both 2D and 3D, one simply chooses a probability $p$ of specular scattering before running the simulation. Then, each time a phonon strikes the surface, the phonon is specularly scattered with probability $p$ and is otherwise diffusely scattered into a randomly chosen direction using Eqs.~ (\ref{equ:3Dlambert}) or Eq.~(\ref{equ:2Dlambert}) for 3D or 2D, respectively.

\begin{figure}
	\includegraphics[width = \columnwidth]{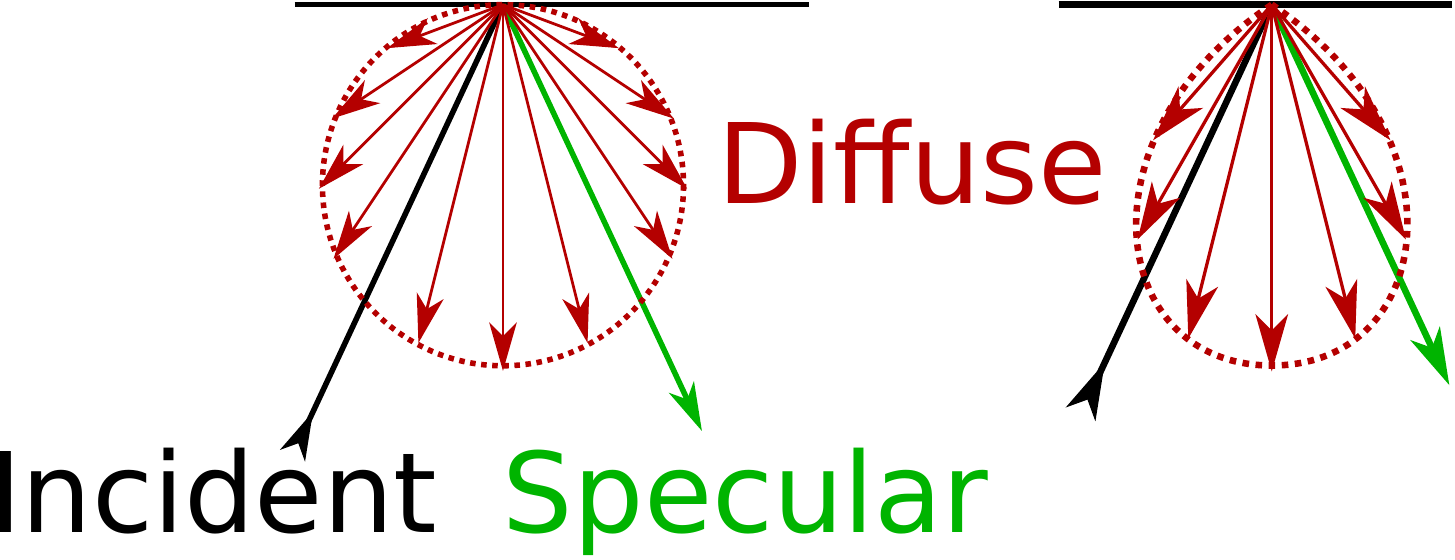}
	\caption{Polar plots of the angular distribution of the outgoing phonon momentum upon diffuse scattering from a surface with (left) a constant specularity parameter and (right) Soffer's momentum-dependent specularity parameter, Eq.~(\ref{equ:soffer}) \cite{Soffer_JAP_67}. Soffer's momentum-dependent specularity parameter decreases the chance of diffuse scattering as $\theta$ (the angle between the phonon wave vector and the surface normal) increases. To satisfy detailed balance, the distribution of outgoing phonons must match the probability of diffuse scattering. The result is a teardrop-shaped distribution, which suppresses scattering at large $\theta$ relative to the constant-specularity-parameter model.}
	\label{fig:boundary_scattering}
\end{figure}

The situation is somewhat more complicated for momentum-dependent specularity-parameter models. Take Soffer's model in 3D \cite{Soffer_JAP_67}. The probability that a phonon scatters specularly is

\begin{equation}\label{equ:soffer}
 p\left(\theta, \phi \right) \propto e^{-\left(2 \sigma \left| \bm{q} \right| \cos{\theta} \right)^2},
\end{equation}

\noindent
where $\sigma$ is the RMS roughness of the surface. Because the probability of diffuse scattering now depends on the angle of incidence, the outgoing phonon distribution no longer follows a simple cosine law (Fig. \ref{fig:boundary_scattering}). We will first attempt to use the inversion technique to find the correct distribution of scattered phonons; after encountering difficulties, we will turn to the rejection technique.

The angular distribution of phonons incident on the surface is still proportional to $\cos{\theta}$, and the probability of diffuse scattering is $1-p\left(\theta, \phi \right)$, so the distribution of incoming phonons that will be diffusely scattered is

\begin{align}
 \begin{split}
  f\left(\theta, \phi \right) &= C \cos{\theta} \left( 1-p\left(\theta, \phi \right) \right) \\
  &= C \cos{\theta} \left( 1- e^{-\left(2 \sigma \left| \bm{q} \right| \cos{\theta} \right)^2} \right),
 \end{split}
\end{align}

\noindent
where $C$ is a normalization constant given by

\begin{align}
 \begin{split}
  1 &= \int_{0}^{\pi/2}\int_{0}^{2\pi} f\left(\theta^\prime,\phi^\prime\right) \sin{\theta^\prime} d\theta^\prime d\phi^\prime, \\
  C^{-1} &= 2 \pi \left(1-\frac{1-e^{-4\sigma^{2}q^{2}}}{4\sigma^{2}q^{2}}\right).
 \end{split}
\end{align}

\noindent Now that $C$ is known, we can find the CDF

\begin{align}
    \begin{split}
        F\left(\theta\right) &= \int_{0}^{\theta}\int_{0}^{2\pi} f\left(\theta^\prime,\phi^\prime\right) \sin{\theta^\prime} d\theta^\prime d\phi^\prime \\
        &= \frac{\frac{e^{-4\sigma^{2}q^{2}}-e^{-4\sigma^{2}q^{2}\cos^{2}\left(\theta\right)}}{4\sigma^{2}q^{2}}+\sin^{2}\left(\theta\right)}{1-\frac{1-e^{-4\sigma^{2}q^{2}}}{4\sigma^{2}q^{2}}}.
    \end{split}
\end{align}

\noindent We have been able to do the integration step analytically, but we cannot do the inversion step analytically. So,

\begin{equation}
 r_\theta= \frac{\frac{e^{-4\sigma^{2}q^{2}}-e^{-4\sigma^{2}q^{2}\cos^{2}\left(\theta\right)}}{4\sigma^{2}q^{2}}+\sin^{2}\left(\theta\right)}{1-\frac{1-e^{-4\sigma^{2}q^{2}}}{4\sigma^{2}q^{2}}}
\end{equation}

\noindent
will have to be solved numerically, for example with Newton's method (which requires computing $\frac{dF}{d\theta}$) or the bisection method.

The other option is to use the rejection technique. This is particularly appealing because we already have a good distribution function $g\left(\theta, \phi \right) \ge f\left(\theta, \phi \right)$, namely $g\left(\theta, \phi \right)=\cos{\theta}$, which is the distribution we use for the constant-specularity parameter model.\footnote{We could also use a constant distribution function for $g\left(\theta, \phi \right)$ as described in Sec. \ref{sec:rejection}, but that would result in reduced performance.} The rejection method to find the outgoing angle $\theta^\prime$ of a scattered phonon is then:

\begin{enumerate}
 \item Generate a random variate $\theta^\prime = \arcsin\left(\sqrt{r_{\theta}}\right)$ where $r_\theta$ is uniformly distributed in $\left[0,1\right]$.
 \item Generate a random variate $y$ that is uniformly distributed in $\left[0, \cos{\theta^\prime} \right]$
 \item If $y < f\left(\theta^\prime, \phi \right) = \cos{\theta^\prime} \left( 1- e^{-\left(2 \sigma \left| \bm{q} \right| \cos{\theta^\prime} \right)^2} \right)$, then use $\theta^\prime$ as your random variate. Otherwise, return to step (1).
\end{enumerate}

\noindent Because we have a good $g\left(\theta, \phi \right)$, and because the inversion method will require an iterative solution, the rejection method is both faster and simpler to implement.

\section{Contacts}\label{sec:contacts}

In the PMC transport simulation, we need contacts at the ends of the simulation domain to act as fixed-temperature phonon reservoirs. These reservoirs must inject new phonons into the simulation domain and absorb the phonons that leave it. There are two basic methods to implement contacts: either the boundaries of the simulation domain mimic reservoirs outside the simulation domain, or parts of the simulation domain get turned into reservoirs. We will call the former a \textit{boundary contact} and the latter an \textit{internal contact}.

Both approaches can make use of existing code, because creating new phonons has algorithmically much in common with scattering existing phonons. Internal contacts can reuse the code for internal scattering mechanisms, and boundary contacts can reuse the code for diffuse surface scattering. All PMC simulations will implement internal scattering mechanisms, so it is always straightforward to implement internal contacts. In contrast, not all PMC simulations will implement diffuse surface scattering, so it may take extra work to implement boundary contacts. We will see that boundary contacts require fewer new phonons to be generated each time step, which increases the performance of the simulation.

\subsection{3D Internal Contacts}\label{sec:3Dinternal}

Internal contacts are relatively simple to implement because the approach almost exclusively reuses code from other parts of the simulation. To implement internal contacts in 3D, one simply deletes all the phonons in a volume near at the end of the simulation domain and then fills the volume with new phonons drawn from the equilibrium distribution (Sec. \ref{sec:3Ddrawing}). The only size requirement is that the volume must be large enough that no phonons can traverse the volume from end to end in one time step.

There are two equivalent ways to fill the volume: either with a certain number of phonons (``fill by number''), or by adding phonons until the cell has the correct energy (``fill by energy''). It is simpler to implement code that fills by number, but the simulation will already need code to fill volumes by energy, in which case filling by energy can simply re-use existing code. We will discuss how to fill by energy in Sec. \ref{sec:conservation}, so we explain how to fill by number here. (Boundary contacts will also use number rather than energy.)

The average number of phonons in branch b in the volume at a temperature $T$ is\footnote{Eq. (\ref{equ:number}) is similar to Eq. (\ref{equ:numberDOS}) except that that latter uses the DOS instead of an integral over reciprocal space.}

\begin{equation}\label{equ:number}
 \mathcal{N}_{\mathrm{b}}(T)=V \int \frac{d^3\bm{q}}{\left(2 \pi\right)^3} \langle n_\mathrm{BE}(\omega_{\mathrm{b},q},T)\rangle,
\end{equation}

\noindent
where the integral is over all allowed $\bm{q}$ (the Brillouin zone in the case of a full dispersion relation). The expression for the average energy in the volume is the same except that a factor of $\hbar \omega_i\left(\bm{q}\right)$ is included in the integral.

The above expression can be simplified for an isotropic dispersion relation:

\begin{align}
    \begin{split}
       \mathcal{N}_{\mathrm{b}}&= V \int_0^{2\pi} \int_0^{\pi} \int_0^{q_\mathrm{max}} \langle n_\mathrm{BE}(\omega_{\mathrm{b},\bm{q}},T)\rangle \frac{q^2 \sin{\theta} dq d\theta d\phi}{\left(2 \pi\right)^3}, \\
                        &= \frac{V}{2 \pi^2} \int_0^{q_\mathrm{max}} \langle n_\mathrm{BE}(\omega_{\mathrm{b},q},T)\rangle q^2 dq,
    \end{split}
\end{align}

\noindent
where $q = \left|\bm{q}\right|$.

$\mathcal{N}_{\mathrm{b}}$ will not be an integer in general. So, add $\lfloor \mathcal{N}_{\mathrm{b}} \rfloor$ phonons and then randomly, with a probability of $\mathcal{N}_{\mathrm{b}} - \lfloor \mathcal{N}_{\mathrm{b}} \rfloor$, add one more phonon. For example, if $\mathcal{N}_{\mathrm{b}} = 1192.63$, then always add 1192 phonons and add an additional phonon 63\% of the time. In this case, it might seem that the additional phonon probability can be done without, but incorrectly generating even one excess phonon per time step per cell may lead to a failure of the simulation (see Sec. \ref{sec:conservation}).

\subsection{3D Boundary Contacts}\label{sec:3Dboundary}

The advantage of boundary contacts is that, instead of having to generate all the phonons inside a volume every time step, you only need to generate the phonons which would have drifted into the simulation domain from a reservoir outside the simulation domain. This effectively makes a boundary contact a blackbody that emits phonons into the simulation domain, so we can use results for the theory of blackbody radiation \cite{Modest_book}. Boundary contacts have much in common with diffuse scattering (Sec. \ref{sec:diffuseBC}).

The expectation number of phonons of branch b and wave vector $\bm{q}$ passing through a surface per unit area and time is

\begin{equation}\label{equ:bb}
    n_\mathrm{b}\left(\bm{q}, T\right) = \left| \bm{v}_{g,\mathrm{b}}(\bm{q}) \right| \cos \theta \langle n_\mathrm{BE}(\omega_{\mathrm{b},\bm{q}},T)\rangle \frac{d^3\bm{q}}{\left(2 \pi\right)^3},
\end{equation}

\noindent
where $\theta$ is the angle between the group velocity $\bm{v}_{g,\mathrm{b}}(\bm{q})$ and the surface normal. Because we only are interested in the number of phonons entering the domain, we only consider $\theta \in \left[0, \pi/2\right]$.

We now turn to the case of an isotropic dispersion relation with $q_\mathrm{max} \ge \left|\bm{q}\right|$ as an upper bound on the wave-vector magnitude. Then $n_\mathrm{b}\left(\bm{q},T\right)$ can be simplified to \cite{Baillis_JHT_08}

\begin{align}\label{equ:3Dbb}
    \begin{split}
        n_\mathrm{b}\left(q, T\right) =& \int_0^{\pi/2}\int_0^{2\pi} v_{g,b}\left(q\right) \cos \theta \langle n_\mathrm{BE}(\omega_{\mathrm{b},\bm{q}},T)\rangle \frac{q^2 \sin \theta dq d\theta d\phi}{\left(2 \pi\right)^3}, \\
        &= v_{g,b}\left(q\right) \langle n_\mathrm{BE}(\omega_{\mathrm{b},\bm{q}},T)\rangle \frac{q^2}{8 \pi^2}dq.
    \end{split}
\end{align}

\noindent Then the total number of phonons entering the simulation domain due to a blackbody of area $A$ in a period of time $\Delta t$ is

\begin{equation}
 \mathcal{N}_{\mathrm{bc}}^{\mathrm{3D}}\left(T\right) = A \Delta t \sum_b \int_0^{q_\mathrm{max}} v_{g,b}\left(q\right) \cos \theta \langle n_\mathrm{BE}(\omega_{\mathrm{b},\bm{q}},T)\rangle \frac{q^2}{8 \pi^2}dq.
\end{equation}

\noindent The cosine in Eq. (\ref{equ:bb}) means that phonons entering the simulation will have the angular distribution from Lambert's cosine law, so we can generate random directions for the incoming phonons using Eq. (\ref{equ:3Dlambert}). In principle, the phonons from the boundary contact will enter the simulation domains at different times, but in practice, all the phonons can be added to the simulation domain at the same time because $\Delta t$ is small.

Then, the overall procedure is to calculate $\mathcal{N}_{\mathrm{bc}}^{\mathrm{3D}}\left(T_h\right)$ and $\mathcal{N}_{\mathrm{bc}}^{\mathrm{3D}}\left(T_c\right)$ for the hot and cold contacts, which are at temperatures $T_h$ and $T_c$ respectively. Then at each time step, any phonons that drift outside the simulation domain are deleted and $\mathcal{N}_{\mathrm{bc}}^{\mathrm{3D}}\left(T_h\right)$ and  $\mathcal{N}_{\mathrm{bc}}^{\mathrm{3D}}\left(T_c\right)$ equilibrium phonons are created (Sec. \ref{sec:3Ddrawing}) at the two contacts, with angular distributions given by Lambert's cosine law, Eq. (\ref{equ:3Dlambert}).

\subsection{2D Contacts}
Two-dimensional internal contacts are essentially the same as their 3D counterpart, except we use the cell area instead of the volume and the integration is over a 2D wave vector $\bm{q}$. When using internal contacts, it does not matter what kind of dispersion we use, because we just initialize the cell to the reservoir energy.

Two-dimensional boundary contacts need more care in implementation. In 2D, to mimic a reservoir from the outside, we need to delete phonons that drift outside of the end-cell boundary (a line) and inject phonons into the cell as if they come from an outside reservoir fixed at temperature $T$. For isotropic dispersion, it is easy to work out the expectation number of phonon to be injected per unit length, per branch, per unit wave number, and per unit time as\footnote{Compare with the 3D equivalent in Eq. (\ref{equ:3Dbb}).}
\begin{equation}
n_\mathrm{b}(q,T)=\frac{1}{4\pi}\langle n_\mathrm{BE}(\omega_{\mathrm{c},j},T)\rangle\frac{q}{2\pi^2}v_{g,\mathrm{b}}(q),
\end{equation}
where $v_{g,\mathrm{b}}(q)$ is the group velocity for phonon with wave number $q$ and branch b. The total expectation number per time step is then
\begin{equation}
\mathcal{N}_{\mathrm{bc}}^{\mathrm{2D}}=H\Delta t\sum\limits_\mathrm{b}\int_{0}^{q_\mathrm{max}}n_\mathrm{b}(q)dq,
\end{equation}
where $\Delta t$ is the time step and $H$ is the height of the simulation domain. Like in 3D, we only generate $\lfloor \mathcal{N}_{\mathrm{bc}}^{\mathrm{2D}}\rfloor$ phonons each step and add an additional phonon with a chance of $(\mathcal{N}_{\mathrm{bc}}^{\mathrm{2D}}-\lfloor \mathcal{N}_{\mathrm{bc}}^{\mathrm{2D}}\rfloor)$. The CDF of angle in 2D is
\begin{equation}
	 F(\theta)=\frac{\int_{0}^{\theta}\cos\theta'd\theta'}{\int_{0}^{\pi}\cos\theta'd\theta'}=\sin\theta.
\end{equation}
The $\cos\theta'$ term comes from the fact that phonons come in at a rate proportional to the group velocity perpendicular to the boundary. Then inversion method can be used to choose the angle as $\theta=\arcsin(r_\theta)$, just like for 2D diffuse scattering, Eq. (\ref{equ:2Dlambert}). When $H$ is small, which is typically the case for PMC for quasi-one-dimensional graphene ribbons, the expectation number of injected particles is too small ($<10$) to represent the distribution, even though the distribution would be correct if we drew a large number of phonons. As a result, the 2D boundary contacts have stability issues and the simulation may take a longer time to converge than the internal-contact implementation.

\section{Energy Conservation}\label{sec:conservation}

If we use the rejection technique to generate a phonon in a cell for PMC simulation, the phonon energy is independent of how much energy is already in the cell. As a result, it is impossible to infuse the cell with exactly the  expected energy. It is tempting to force the last phonon's energy to achieve exact energy conservation. However, by doing so, we are introducing excess phonons with energies smaller than average, i.e., more of them than the equilibrium distribution would predict. Owing to the low energies, these phonons have large group velocities and small scattering rates, so they carry their energy out of the simulation domain almost ballistically and lead to energy depletion inside the structure. Even generating a single phonon from the wrong distribution in every cell once a time step can add up and cause the simulation to fail dramatically.

To respect the phonon energy distribution, we allow the energy inside a cell to be uncertain up to half the maximal energy of a single phonon: we consider the desired cell energy to be reached with satisfactory accuracy if the cell energy falls within  $\left[E-\frac{\hbar\omega_\mathrm{max}}{2},E+\frac{\hbar\omega_\mathrm{max}}{2}\right]$, where $E$ is the exact desired cell energy and $\hbar\omega_\mathrm{max}$ is the maximum energy carried by a phonon in the material.

Furthermore, the treatment of inelastic scattering does not conserve energy precisely at each time step, but only in an average sense \cite{Knezevic_JAP_14}. To better conserve the energy after scattering, each cell is reinitialized after each time step by adding/deleting phonons from the equilibrium distribution. The accumulation of offset energy might cause a problem, so we record the offset energy $E_\mathrm{offset}=E-E_\mathrm{actual}$ at each step and add it to the desired energy at the next reinitialization.

During drift, phonons might cross cell boundaries and therefore change the total energy in each cell. The energy of the phonons that drifted in (out) has to be added (subtracted) before entering the scattering routine. As a result, we record the cell energy just after the phonons finish drift in a prescattering array, \(E_\mathrm{prescat}\). Together with the array \(E_\mathrm{i,offset}\), which records the offset energy from our last attempt at enforcing energy conservation, we calculate the desired energy after scattering as \(E_{\mathrm{d}}=E_{\mathrm{prescat}}+E_{\mathrm{offset}}\). Then we can scatter the phonons and calculate the actual cell energy after scattering, stored in an after-scattering array, \(E_{i,\mathrm{afterscat}}\). Again, we only enforce that the cell energy $E$ get into the range \( [E_{\mathrm{d}}-\frac{\hbar\omega_\mathrm{max}}{2}, E_{\mathrm{d}}+\frac{\hbar\omega_\mathrm{max}}{2}]\), where $E_{\mathrm{d}}$ is the desired energy. The initial cell energy in the reinitialization process is  $E_i=E_{i,\mathrm{afterscat}}$. For each cell $i$, we compare $E_i$ and $E_{i,\mathrm{d}}$.
\begin{enumerate}
	\item If $E_i\in [E_{i,\mathrm{d}}-\frac{\hbar\omega_\mathrm{max}}{2}, E_{i,\mathrm{d}}+\frac{\hbar\omega_\mathrm{max}}{2}]$, we consider this cell good and move on to the next one.
	\item If $E_i<E_{i,\mathrm{d}}-\frac{\hbar\omega_\mathrm{max}}{2}$, we generate a phonon from the equilibrium distribution and add it to a random place in the cell. The new cell energy after addition is $E^{\mathrm{new}}_i=E^{\mathrm{old}}_i+\hbar\omega_0$, where $\hbar\omega_0$ is the energy carried by the added phonon. Keep adding phonons until the final $E_i$ falls in the appropriate range.
	\item If $E_i>E_{i,\mathrm{d}}+\frac{\hbar\omega_\mathrm{max}}{2}$, we would randomly choose one phonon in the cell and delete it. \footnote{It is very important that the phonon be chosen randomly. For example, simply deleting the oldest phonon will slowly skew the phonon distribution, because the oldest phonon is likely to be a phonon with a low scattering rate.\\
Depending on the data structure used for storing phonons, deleting algorithms may vary. One can use a random number to choose the index of phonon to be deleted if the structure supports easy random access. For structures like the linked list, one can use the Fisher-Yates shuffle algorithm to generate a random permutation of the array and delete in sequence.} After the deletion, the new energy in the cell is $E^{\mathrm{new}}_i=E^{\mathrm{old}}_i-\hbar\omega_0$ where $\hbar\omega_0$ is the energy carried by the deleted phonon. Then we compare $E^{\mathrm{new}}_i$ and $E_{i,\mathrm{d}}$ again and keep this random deletion until $E^{\mathrm{new}}_i$ falls in the desired range.
\end{enumerate}
After all cells have energies in the desired range, we record the new offset energy as \(E_{\mathrm{offset}}=E_{\mathrm{d}}-E_i\) and use it in the next reinitialization process (after the next time step).

\section{Conclusion}
PMC is a versatile stochasic technique for solving the Boltzmann equation for phonons in structures that can have real-space roughness and experimentally relevant sizes. We presented the relative merits of inversion vs rejection techniques for generating random variates to represent the random variables with nonuniform distributions, which are relevant in thermal transport: generating the attributes for phonons in equilibrium with full and isotropic dispersions, randomizing outgoing momentum upon diffuse boundary scattering, implementing contacts (boundary and internal), and conserving energy in the simulation. We also identified common themes in phonon generation and scattering that are helpful for reusing code in the simulation (generating thermal phonon attributes vs internal contacts, diffuse surface scattering vs boundary contacts). We hope these examples will inform the reader about both the mechanics of random variate generation and choosing a good approach for whatever problem is at hand, and aid in the more widespread use of PMC for thermal transport simulation.

\section{Acknowledgement}
The authors gratefully acknowledge support by the U.S. Department of Energy, Office of Basic Energy Sciences, Division of Materials Sciences and Engineering under Award DE-SC0008712. This work was performed using the compute resources and assistance of the UW-Madison Center for High Throughput Computing (CHTC) in the
Department of Computer Sciences.

\bibliographystyle{psp-book-har}    
\bibliography{references}      


\printindex

\end{document}